\documentclass[aps,prl,preprint]{revtex4-1}
\usepackage{graphicx}
\usepackage{epstopdf}
\begin{document}

\title{Capillary force-induced structural instability in liquid infiltrated
elastic circular tubes}

\author{Y. Yang$^1$, Y. F. Gao$^1$, D. Y. Sun$^1$, M. Asta$^{2,3}$ and
J. J. Hoyt$^4$ }
\address{$^1$Department of Physics, East China Normal University,
Shanghai 200062, P. R. China}
\address{$^2$Department of Materials Science and Engineering, University of
California, Berkeley, CA 94720}
\address{$^3$Department of Chemical Engineering and Materials Science,
University of California, Davis, CA 95616}
\address{$^4$Department of Materials Science and Engineering, McMaster
University, Hamilton, ON, Canada}

\date{\today}

\begin{abstract}
The capillary-induced structural instability of an elastic circular
tube partially filled by a liquid is studied by combining
theoretical analysis and molecular dynamics simulations. The
analysis shows that, associated with the instability, there is a
well-defined length scale (elasto-capillary length), which exhibits
a scaling relationship with the characteristic length of the tube,
regardless of the interaction details. We validate this scaling
relationship for a carbon nanotube partially filled by liquid iron.
The capillary-induced structural transformation could have potential
applications for nano-devices.
\end{abstract}
\maketitle

A wide range of phenomena in nature, which span from everyday
observations to many bio-related processes, are a result of
capillary forces. Common examples include the shape of liquid
droplets, the imbibition of a sponge \cite{deGennes03}, the clumping
of wet hair into bundles \cite{Bico04} and coalescence of paintbrush
fibers \cite{Kim06}, the standing of aquatic insects on water
\cite{Hu05}, and lung airway closure \emph{etc} \cite{Heil08}.
Recently it was found that capillary forces can induce structural
deformations or instability in an elastic system, in cases where
these surface forces are comparable in magnitude to bulk elastic
forces. More importantly, there is a well-defined length scale
(elasto-capillary length $L_{EC}$) which underlies such structural
instabilities \cite{Cohen03}, which usually reveals an intrinsic
scaling relation with the characteristic dimensions
\cite{Cohen03,Bico04,Kim06,py07,Huang07,Neukirch07}. The
elasto-capillarity deformation or instability has been also found to
have many interesting applications
\cite{Chakrapani04,py07,Huang07,Pokroy09}.

Liquid infiltrated elastic tubes are common in bio-systems and
everyday life. When the capillary forces are comparable to the
bending stiffness of the tube, the tube can be deformed or become
unstable at some critical filling fraction. Whether the instability
can be associated with a new type of scaling relationship has thus
far remained relatively unexplored. In this work, molecular dynamics
(MD) simulations of this phenomenon are coupled with theoretical
analysis to investigate this issue for an elastic cylindrical tube
partially filled by a liquid. Our results show that the instability
or deformation of a flexible tube leads to a well-defined scaling
law regardless of the interaction details. Our MD simulations on
liquid iron encapsulated by carbon nanotubes quantitatively
demonstrate the scaling law.

We consider an elastic circular tube with length of $L_{t}$ and
radius of $R_{0}$ partially filled with an incompressible Newtonian
liquid of length $L_l$, where $R_{0}$ refers the radius of the ``dry" tube  without any compression.
The upper panel of Fig. 1 illustrates the model used in the
current work. The energy of the system can be written formally as,
\begin{eqnarray}
E=E_{b}+E_{c}+E_{i} \label{YYeq1}
\end{eqnarray}
where $E_{b}$, $E_{c}$ and $E_{i}$ are the elastic bending strain
energy, elastic compression strain energy, and interfacial free
energy (capillary energy), respectively. In the continuum elasticity
picture \cite{Landau96,Timoshenko88}, the bending strain energy is
$E_{b}=\frac{L_{t}B}{2}\oint\frac{1}{\rho^2}dl$, and the compression
strain energy is $E_{c}=\frac{L_{t}C}{ 2}\oint(\frac{\oint
dl-P_0}{P_0})^2dl$, where $\oint$ denotes the curvilinear integral
along the perimeter of the cross section and $P_0=2\pi R_0$ is the
perimeter of the original tube cross section. The bending stiffness
$B=Yh^3/12(1-\nu^2)$ and the compression stiffness $C=Yh/(1-\nu^2)$
are constants related to Young's modulus $Y$, the Poisson ratio
$\nu$, and the tube thickness $h$. The local radius of curvature is
denoted as $\rho$. The interfacial free energy is, $E_{i} =
A_{lv}\gamma_{lv} +A_{sl}\gamma_{sl}+A_{sv}\gamma_{sv}+2P\tau$,
where $A_{lv}$, $A_{sl}$ and $A_{sv}$ are the areas of liquid-vapor,
solid-liquid and solid-vapor interfaces, respectively, $P$ is the
perimeter of the cross section and $A_{sv}=2PL_{t}-A_{sl}$, here we
have included the solid-vapor surface outside the cylinder that is
not covered by liquid. The surface and interface free energies are
denoted $\gamma_{lv}$, $\gamma_{sl}$ and $\gamma_{sv}$, for the
liquid-vapor, solid-liquid and solid-vapor interfaces, respectively.
The specific free energy of the three-phase contact line (i.e., the
line tension) is denoted as $\tau$.\cite{Boruvka77} The $2P\tau$
term usually can be neglected if the dimension of the system is not
too small.

Since only near-critical sizes (where the circular shape is near its
stability limit) are considered, the non-circular shape is
approximated by that of an ellipse. For an ellipse with long axis
$R_a$ and short axis $R_b$, the shape can be described well by
introducing two independent variables, $R$ and $\omega$, which are
defined as $R=\sqrt{R_aR_b}$ and $\omega=R_a/R_b$. When $\omega$=1,
these variables describe a perfect circle.  By introducing two integrals,
$f_{1,2}(\omega)=\frac{1}{2\pi}\oint(\omega\cos^{2}t+
\omega^{-1}\sin^{2}t)^{\alpha_{1,2}}dt,$ with $\alpha_1=-2.5$ and
$\alpha_2=0.5$, we can write $\oint\frac{1}{\rho^2}dl=\frac{2\pi}{R}f_1(\omega)$
 and $\oint dl=P=2\pi Rf_2(\omega).$\cite{Sun04} Finally we have
\begin{eqnarray}
E_{b}+E_{c}=\frac{B\pi L_{t}}{R}f_1(\omega)+C\pi
R_0L_{t}[(\frac{R}{R_0}f_2(\omega))^3-2(\frac{R}{R_0}f_2(\omega))^2+\frac{R}{
R_0}f_2(\omega)].\label{YYeq2}
\end{eqnarray}

In order to calculate the capillary energy, the area of the interfaces
should be written in explicit forms. However, the general calculation of
the liquid-vapor interface, which is a meniscus, for arbitrary tube shape
is beyond the scope of the present treatment and will be discussed in a
subsequent paper \cite{Yang09}. To elucidate the important physics as well
as to simplify the mathematical treatment, we consider the cases where
the contact angles are close to $90^o$. Within this limitation, all
the interface areas can be defined without ambiguity.  It should be
noted that in the simulations discussed below, the solid-liquid contact
angles are in fact approximately $90^o$.

Thermal fluctuations of the tubes are neglected as they exist on a
length scale comparable to the persistence length($l_p=YI/k_BT$,
and Y: Young's modulus, I: the area moment of intertia. ), which usually is
much longer than the characteristic length of system. For example, in single-walled carbon
nanotubes, the persistence length is on the order of 45 $\mu m$
\cite{Cohen03}. With the above assumptions as well as the assumption
of that the liquid is incompressible, we can write $A_{lv}=2\pi
R^2$, $A_{sl}=PL_l(\frac{R_0^2}{R^2})$ and $A_{sv}=2PL_t-A_{sl}$.
So, the interfacial free energy is,
\begin{equation}
E_{i}=2\pi R^2\gamma_{lv} +PL_{l}\frac{R_0^2}{R^2}(\gamma_{sl}-\gamma_{sv}) +
 2PL_{t}\gamma_{sv} + 2P\tau \label{YYeq3}
\end{equation}

In Eq. 2, the elastic energy always tends to maintain the tube in a
circular shape. If the capillary energy approaches or exceeds the
magnitude of the elastic energy, the circular shape will be unstable
and the system might seek a lower energy state. Physically, the
deformation of the tube reduces the interfacial energy at the cost
of increasing the elastic energy. For $\theta<90^o$,
$(\gamma_{sl}-\gamma_{sv})<0$, the system could deform to increase
the area of the solid-liquid interface. Even for $\theta\geq90^o$,
$(\gamma_{sl}-\gamma_{sv})\geq0$, the tube still has a chance to
deform if the first term exceeds the second term in Eq. 3, in which
case the area of the liquid-vapor interface decreases. The MD
simulations in the current work focus on the latter case.
The critical transition point can be located by the condition
$\frac{\partial^2E}{\partial\omega^2}|_{\omega=1}=0$. And
the radius of tube at the critical point ($R_c$) is determined by $\frac{\partial E}{\partial R}|_{\omega=1}=0$.
Remembering $f_{1,2}(\omega)|_{\omega=1}=1$,
$f'_{1,2}(\omega)|_{\omega=1}=0$ and
$f''_{1,2}(\omega)|_{\omega=1}=\frac{15}{8},\frac{3}{8},$ the above
conditions produce the critical relation,
\begin{eqnarray}
L_{t}^{\ast}&=&\frac{2[\gamma_{lv}-(\gamma_{sl}-\gamma_{sv})\cdot\frac{L_{l}R_0^2}{R_c^3}]\cdot R_c^3}{3B} \label{YYeq5}
\end{eqnarray}
where $L_{t}^{\ast}$ is the critical length of the tube.
For elastic tubes, if the compression stiffness ($C$) is much larger than the bending stiffness $B$,
which is the the case we are interested in, one can neglect the different between $R_0$ and $R_c$,
\begin{eqnarray}
L_{t}^{\ast}&=&\frac{2[\gamma_{lv}-(\gamma_{sl}-\gamma_{sv})\cdot\frac{L_{l}}{
R_0}]\cdot
R_0^3}{3B} \label{YYeq5}
\end{eqnarray}

We can define an effective interface free energy
$\gamma^*=[\gamma_{lv}-(\gamma_{sl}-\gamma_{sv})\cdot
\frac{L_{l}}{R_0}]$, where the factor $\frac{L_{l}}{R_0}$ in
$\gamma^*$ reflects an inherent geometric relation among the three
interfaces. Note that if $2\gamma^*$ is divided $L_{t}^{\ast}$,
according to Eq. 5, a generalized pressure
$p^*=\frac{2\gamma^*}{L_t^{\ast}}=\frac{3B}{R_0^3}$ can be obtained,
which is the same form as reported in Ref.~\cite{Sun04} for CNTs
under pressure. In fact, this capillary force can be regarded as a
kind of negative internal pressure. From Eq. 5, a generalized
elasto-capillary scaling relation can be derived.
\begin{equation}
\left[ \frac{2R_0^3}{3L_{t}^{\ast}} \right]^{\frac{1}{2}}=
\left[ \frac{B}{\gamma^*} \right]^{\frac{1}{2}} \label{YYeq6}
\end{equation}
Similar to previous studies,
\cite{Cohen03,Bico04,Kim06,py07,Huang07,Neukirch07} we can define
an elasto-capillary length ($L_{EC}=[\frac{B}{\gamma^*}]^{\frac{1}{
2}}$) for this system, which gives the typical effective curvature
induced by capillarity on the tube. Different from previous studies,
the elasto-capillary length is not simply defined in terms of
interfacial free energies, but rather an effective (average) interface
free energy ($\gamma^*$). The difference stems from the fact that in
the filled liquid tube system studied here, the area of the
solid-liquid (solid-vapor) and liquid-vapor interfaces changes in
different ways (amounts) as the tube is deformed. The former
scales as \emph{R}, while the later scales as $R^2$. It implies that
the average capillary forces depend on the system size itself
($\frac{L_{l}}{R_0}$ in $\gamma^*$ reflects the dependence). However
in previous studies, either the three interfaces change
simultaneously by the same amount (liquid drop on thin
films)\cite{py07,Huang07} or only one interface changes area
(e.g., slender rods immersed in liquid)
\cite{Cohen03,Bico04,Kim06,Neukirch07}, and thus the average
capillary force is independent on the size of system.

We can also define a characteristic length of the system at the
critical point, $L_{C}=\frac{\emph{Area of cross
section}}{\sqrt{\emph{Surface area of tube}}}=\frac{2\pi
R_0^2}{\sqrt{2\pi R_0L_{t}^{\ast}}}=\sqrt{3\pi}(\frac{
2R_0^3}{3L_{t}^{\ast}})^{\frac{1}{2}}$, which is a purely geometric
description of the tube. Eq. 6 becomes a generic scaling
relationship,
\begin{equation}
L_{C}=\sqrt{3\pi}L_{EC} \label{YYeq7}
\end{equation}
$L_C$ defines a typical length scale at which the elastic energy and
surface energy  are comparable. Different from previous studies for
slender rods immersed in liquid,
\cite{Cohen03,Bico04,Kim06,Neukirch07} or for liquid drops on thin
films,\cite{py07,Huang07} the characteristic length in the present
case is not simply a spatial dimension of system. The reason is that
the elastic energy depends on both $R_0$ and $L_t$, where $R_0$ and
$L_t$ are functions of the radial bulk modulus and total elastic
energy respectively. Therefore, the elasto-capillary length and
the scaling relationship presented here represent a new type of
instability process.

To test the theoretical analysis, we simulate liquid iron
encapsulated by a single-walled carbon nanotube (SWCNT) using MD
simulation. All the simulations made use of the LAMMPS (large-scale
atomic/molecular massively parallel simulator) code
\cite{Plimpton95}. The interaction between carbon atoms is described
by the second-generation reactive empirical bond order (REBO2)
potential.\cite{Brenner02} The many-body potential developed by
Mendelev \emph{et al} is used to describe the Fe-Fe interactions.
\cite{Mendelev03} The weak interaction between C and Fe is
modeled by a truncated Lennard-Jones(LJ) potential
\cite{Broughton83,Davidchack03}. The interaction parameters $\sigma$
and $\epsilon$ are obtained by fitting to ab-initio
results,\cite{Durgun03} $\sigma=2.05\AA$ and$\epsilon=0.09463eV$.
The ab-initio results are also used by other authors to fit the C-Fe
interaction based on Johnson pair form potential.\cite{Ding04}
This fitted potential give a reasonable contact angle ($\sim 107^o$
at 2500K) compared to the experimental value ($\sim 125^o$ at 923K)
for liquid Fe in SWCNT \cite{Wei07}.

In the present study, periodic boundary conditions in the axial
direction and free boundary conditions in the radial directions are
adopted. The systems are simulated at T=2500 K and zero pressure.
The temperature is much higher than the melting point of Fe
,\cite{Mendelev03} so that Fe is in a well defined liquid state. A
typical run lasts 3ns for each case. The tubes used in this work are
of the zigzag type, including (32,0),(40,0),(46,0) and (50,0), with the
typical length in the range of 30${\AA}$ to 250${\AA}$. To locate
the critical length for each sample, the length of the tube is
decreased until the circular shape of the CNT becomes unstable.

To quantitatively compare the model prediction and MD results, the
value of all the variables on the right hand side of Eq. 5 are
needed.  $\gamma_{lv}$ for
liquid iron at 2500K is calculated separately using the method
reported in Refs.~\cite{Nijmeijer88,Sides99}, the value obtained is
about 0.82$\pm$0.02$J/m^2$. Both ($\gamma_{sl}-\gamma_{sv}$) and
$\tau$ can be calculated from the extended Young's equation
\cite{Boruvka77} if the contact angle as a function of the radius of
the CNT is known. Contact angles $\theta$ are measured during MD
simulations for a circular shape. The determination of the contact
angle follows the procedure described in
Refs.\cite{Kutana07,Werder03}. We obtain
$\gamma_{sl}-\gamma_{sv}$=0.24$\pm$0.02$J/m^2$ and
$\tau=0.4\pm0.3\times10^{-10}J/m$. The elastic constant $B$ of the
CNT for the REBO2 potential is adopted from Ref.~\cite{Lu09},
$B$=1.4$eV$.

The measurement of the CNT shape is based on the calculation of the
cross sectional area $S$. An order parameter thus can be defined as
$\lambda=<S>_t/S_0$, where $S_0$ is the initial cross sectional area
of the CNT and $<...>_t$ denotes the time average. With this definition
$\lambda$ is one for the
circular shape while it is smaller than one for a non-circular shape,
and decreases continuously with increasing deviation from a circular
shape. Due to thermal fluctuations, the shape of the CNT varies with
time. We define the critical length for which $\lambda$ is smaller
than one within the statistical uncertainty.

The lower panel of Fig. 1 shows the time-averaged cross section for
a (40,0) CNT filled with 16.85${\AA}$ liquid for a few selected
$L_t$. With decreasing $L_t$, the shape of the cross section changes
from a circle to an ellipse. Fig. 2 shows the evolution of $\lambda$
vs the length of the CNT ($L_t$) for a (40,0) tube, where the
different symbols denote systems with different lengths of liquid.
We find that the order parameter can clearly distinguish the CNT
shape change from circle to ellipse. From Fig. 2, one can see that,
for longer $L_t$, $\lambda$ remains at an average value of one
(corresponding to a circular shape) within the error bars, while
$\lambda$ deviates from one for shorter $L_t$. The critical length
can be located from this figure. It should be pointed out that the
critical length is not dependent on the specific definition of order
parameter. Another order parameter based on the Fourier spectrum of
a shape fluctuation function \cite{Khare96,Schlober99} was also
tested and the results do not show significant differences. For all
studied samples, when $L_t$ is smaller than a certain critical
length, the shape of the cross section will change from circle to
ellipse. The shape changes are purely elastic deformations and the
system can relax reversibly by removing the liquid.

Fig. 3 shows the critical length of the tube $L_t^\ast$ vs the
liquid length $L_l$ for all CNTs studied, where the solid line is
plotted based on Eq. 5 using the parameter values given above. One
can see that the MD results are in excellent agreement with the
theoretical predictions. The MD results are consistent with a linear
relationship between $L_{C}$ and $L_{EC}$ as expected from the model
analysis (see Fig. 4). A best fit produces a slope of
3.034$\pm$0.014, which is in excellent agreement with the
theoretical prediction of $\sqrt{3\pi}$.

The critical length (Eq. 5) affords a few potential applications in
nano-science or in designing nano-devices. One possible application
is the measurement of the elastic bending stiffness ($B$) of the
tube or the contact angle. For this purpose, we only need knowledge
of $L_t^\ast$ as a function of $L_l$ by filling liquid into the
tubes. If the $\gamma_{lv}$ and contact angle are known, $B$ can be
obtained. On the other hand, if $B$ is known, the contact angle can
be estimated. The capillary induced structural transformation also
provides an ideal mechanism of conversion of capillary energy to
elastic energy. If the values of the interface energies in Eq. 5 are
sensitive to the temperature, or to other environment variables, an
environment-controlled nano-machine could be achieved.

In summary, the capillary-induced instability of elastic circular
tubes has been studied by combining theoretical analysis and MD
simulation. The instability depends on two key length-scales, the
elasto-capillary length $L_{EC}$ which represents the typical radius
of curvature produced by capillary forces on an elastic tube and the
characteristic geometric length $L_C$, which defines a typical
reduced size of the tube. At the critical point, $L_C$ is found to
be proportional to $L_{EC}$, above which a tube preserves a circular
shape. The MD results are found to be in excellent agreement with
the theoretical prediction. The current results can be useful to the
design of nano-devices.

\begin{acknowledgments}
We thank Prof. B. B. Laird for his helpful discussion. This research
is supported by the National Science Foundation of China, the
special funds for major state basic research and CAS projects.
\end{acknowledgments}

\bibliography{YY}

\begin{figure}
\includegraphics[scale=1.0]{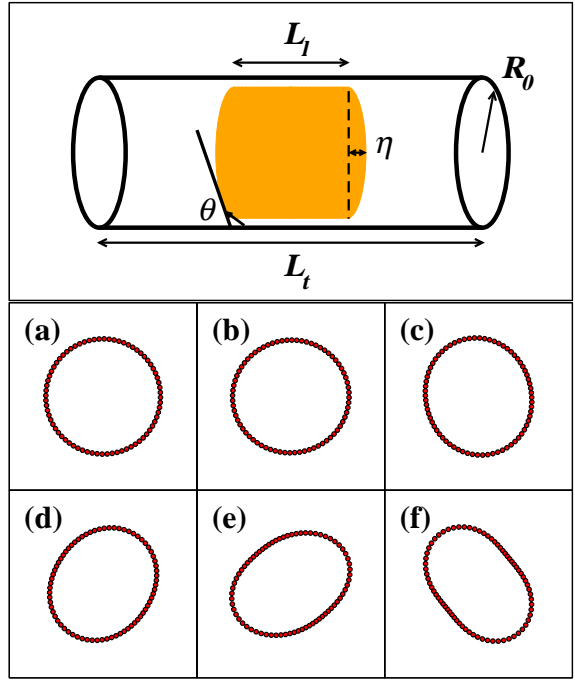}
\caption{\label{figure1} (Color online) Upper panel: Sketch of the
model used in present work. The elastic tube is
characterized by length ($L_t$) and radius ($R_0$). The encapsulated
liquid is described by the length ($L_l$), the height of cap ($\eta$) and
the contact angle ($\theta$). Lower panel: Time averaged cross
section for various lengths of the CNT $L_t$. All results are for
(40,0) CNTs with liquid length $L_l=16.85 \AA$. (a) $L_t=96.92 \AA$,
(b) $L_t=84.28 \AA$, (c) $L_t=59.00 \AA$, (d) $L_t=50.58 \AA$, (e)
$L_t=46.36 \AA$, (f) $L_t=42.15 \AA$.}
\end{figure}

\begin{figure}
\includegraphics[scale=1.0]{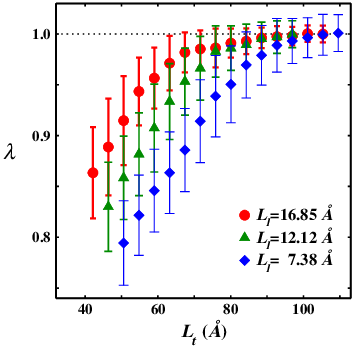}
\caption{\label{figure2} (Color online) The order parameter
$\lambda$ versus the length of the CNT (40,0). Circle: length of liquid
$L_l$=16.85$\AA$, Triangle: length of liquid $L_l$=12.12$\AA$,
Diamond: length of liquid $L_l$=7.38$\AA$.}
\end{figure}

\begin{figure}
\includegraphics[scale=1.0]{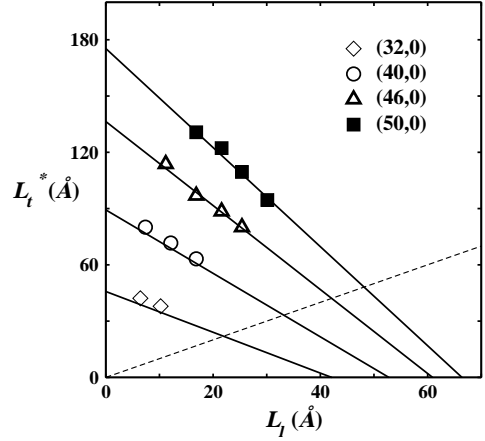}
\caption{\label{figure3}The critical length of the tube as a
function of the length of liquid filled region. Symbols: the
molecular dynamics results, solid lines: the analysis results of Eq.
5, dashed lines: The tube is fully filled with liquid, below this
line the model is not valid.}
\end{figure}

\begin{figure}
\includegraphics[scale=1.0]{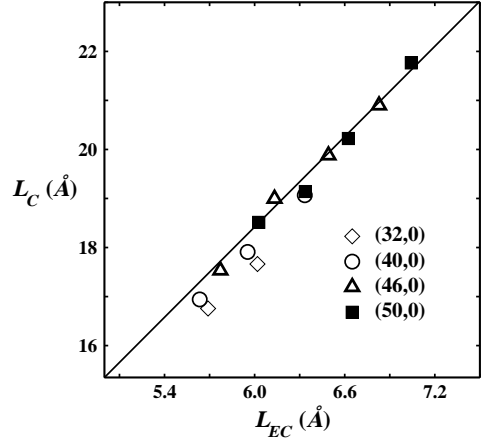}
\caption{\label{figure4}The characteristic geometry length as a
function of elasto-capillary length. Symbols: the molecular dynamics
results, lines: analysis results of Eq. 7.}
\end{figure}
\end{document}